\documentclass[conference]{IEEEtran}
\IEEEoverridecommandlockouts
\usepackage[noadjust]{cite}
\usepackage{amsmath,amssymb,amsfonts}
\usepackage{algorithmic}
\usepackage{graphicx}
\usepackage{textcomp}
\usepackage{xcolor}
\def\BibTeX{{\rm B\kern-.05em{\sc i\kern-.025em b}\kern-.08em
    T\kern-.1667em\lower.7ex\hbox{E}\kern-.125emX}}
\begin{document}

\title{List Autoencoder: Towards Deep Learning Based Reliable Transmission Over Noisy Channels\\
}

\author{\IEEEauthorblockN{Hamid Saber}
\IEEEauthorblockA{\textit{Samsung Semiconductor Inc} \\
\textit{San Diego, USA}\\
hamid.saber@samsung.com}
\and
\IEEEauthorblockN{Homayoon Hatami}
\IEEEauthorblockA{\textit{Samsung Semiconductor Inc} \\
\textit{San Diego, USA}\\
h.hatami@samsung.com}
\and
\IEEEauthorblockN{Jung Hyun Bae}
\IEEEauthorblockA{\textit{Samsung Semiconductor Inc} \\
\textit{San Diego, USA}\\
jung.b@samsung.com}
}

\maketitle

\begin{abstract}
In this paper, we present list autoencoder (listAE) to mimic list decoding used in classical coding theory. With listAE, the decoder network outputs a list of decoded message word candidates. To train the listAE, a genie is assumed to be available at the output of the decoder. A specific loss function is proposed to optimize the performance of a genie-aided (GA) list decoding. The listAE is a general framework and can be used with any AE architecture. We propose a specific architecture, referred to as incremental-redundancy AE (IR-AE), which decodes the received word on a sequence of component codes with non-increasing rates. Then, the listAE is trained and evaluated with both IR-AE and Turbo-AE. Finally, we employ cyclic redundancy check (CRC) codes to replace the genie at the decoder output and obtain a CRC aided (CA) list decoder. Our simulation results show that the IR-AE under CA list decoding demonstrates meaningful coding gain over Turbo-AE and polar code at low block error rates range.
\end{abstract}

\begin{IEEEkeywords}
Auto-encoder, Deep learning, List decoding
\end{IEEEkeywords}

\section{Introduction}

Reliable transmission over noisy channels has been an active research area for the past decades. Channel coding is the main tool to achieve reliable transmission by finding higher dimensional representations of the input data. In his seminal work \cite{Shannon}, Shannon proved the existence of capacity-achieving sequence of codes by random construction of an ensemble and investigating the conditions for the feasibility of reliable communication. Design of channel codes that approach or achieve the channel capacity has since then been an elusive goal. Among most landmark codes designed thus far are Turbo, LDPC and polar codes \cite{TurboCode}-\cite{PolarArikan}.

Traditionally, an $(N,K)$ channel code is constructed by designing an encoder that maps a set of $2^K$ binary message words of length $K$ to a set of $2^K$ codewords of length $N$ for transmission over the  channel. Typically, mathematical analysis is used to tailor the encoder and decoder to one another. For instance, under the maximum a posteriori (MAP) decoder which minimizes the block error rate (BLER), the encoder is designed such that the pairwise distance properties of the code is optimized. MAP decoding is almost never used unless when the code is very short, or can be described via a trellis diagram with rather a small size, similar to convolutional codes. It is also possible to design the decoder first. Polar code design typically follows this approach, where the encoder design, suitably defined in \cite{PolarArikan}, is carried out to optimize the performance under a successive cancellation (SC) decoder. Details aside, almost all  classical code design approaches heavily rely on an information-theoretically well-defined channel model, which in most cases is additive white Gaussian noise (AWGN) channel, and employing mathematical analysis as an essential tool. More importantly, the code design progress has thus far been sporadic and heavily relying on the ingenuity of humans.

There has been a growing interest in automating the design of encoder and decoders using deep learning framework. A deep-learning based framework allows for design of encoder and decoder for channels that cannot be described by a well-defined model or can be described but the model is too complex for code design. Although the ultimate goal of deep learning based design is envisioned to be for arbitrary channels, a first step towards this end can be designing codes which can compete with the state-of-the-art classical channel codes over the AWGN channel. The code design essentially can be applied to any channel provided that a sufficient number of transmissions over the channel are performed to construct a sufficiently large training set.  

Deep learning has been employed to design decoders for the classical encoders \cite{decoder1}-\cite{decoder5}. It has also been used to design both encoder and decoder based on autoencoders (AEs). AEs are powerful deep learning frameworks with a wide variety of applications which fall into two categories: under-complete and over-complete AEs. An under-complete AE is used to learn latent representation of the input data by transforming it to a smaller latent space. Under-complete AEs are used for numerous tasks including  denoising, generative models and representation learning \cite{Denoising}-\cite{representationlearning2}. On the other hand, over-complete AEs add redundancy to the input data to transform it to a higher dimension. One of the main application of over-complete AEs is that the higher-dimensional representation can be transmitted over a noisy channel allowing the receiver to reliably decode the input data \cite{AE1}-\cite{AE7turboae}. In particular, the authors in \cite{AE7turboae} used convolutional neural network (CNN) and recurrent neural network (RNN) to mimic the architecture of classical turbo encoder and decoder. The proposed Turbo-AE is reported to have competitive performance to the state-of-the-art classical codes while being trainable for an arbitrary channel model.

This paper is a further attempt to improve the design of AEs for reliable communication over noisy channels and is based on the concept of list decoding in the classical coding theory. Our contributions are:

\begin{itemize}
\item We present list autoencoder (listAE) as a general deep learning framework applicable to any AE architecture. With listAE, the decoder network outputs a list of decoded message word candidates. The listAE mimics the list decoding in classical coding theory.

\item We provide a specific loss function which operates on the output list. The loss function aims to optimize the performance of a genie-aided (GA) decoder. We assume a genie is available at the decoder output and, whenever the transmitted message word is present in the list, it informs us which candidate it is. In other words, with the GA decoder, a block error event is counted if and only if the transmitted word is not present in the output list. During the testing phase, the functionality of the genie is emulated by using cyclic redundancy check (CRC) code. CRC is appended to the message word prior to encoding by the encoder network. At the decoder, CRC check is carried out for each output candidate to select a single candidate as the final output of the decoder. The concept of CRC-aided (CA) list decoding in widely used in classical coding theory. 

\item listAE can be applied to any AE architecture. With the promising performance of Turbo-AE, it is natural to use its architecture in the listAE framework. While we train and evaluate the performance of listAE with this architecture, we also propose a more general architecture that decodes the received word on a sequence of component codes with non-increasing rates. The architecture, referred to as incremental redundancy AE (IR-AE), illustrates improvement over Turbo-AE architecture for smaller list sizes while having comparable performance at large list sizes. 

\end{itemize}

\section{Problem definition}

\begin{figure}[tb]
\begin{center}
\includegraphics[scale=.48,viewport=15 82 570 220,clip]{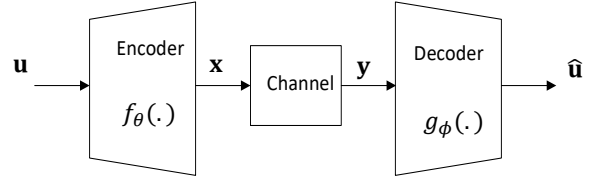}
\end{center}
\vspace*{-4mm}
\caption{Channel coding as an auto-encoder.}
\label{generalAE}
\end{figure}

The problem of reliable transmission over a noisy channel can be defined as follows. As can be seen in Fig. \ref{generalAE}, a message word of $K$ bits is formed as $\mathbf{u}=[u_1,\ldots,u_K]$, where the $u_i$ take binary values from $\{0,1\}$. The message word is encoded using an encoder neural network with an encoding function $f_\theta (.)$ to obtain real-valued codeword $\mathbf{x}=[x_1,\ldots,x_N]=f_\theta(\mathbf{u})$ where the $\theta$ denotes the weights of the encoder neural network and $N$ denotes the code length. A power normalization block is applied to $\mathbf{x}$ to give a codeword with zero mean and unit variance code symbols, i.e. $E(x_i)=0$ and $E(x_i^2)=1$ for $i =1,\ldots,N$. The codeword $\mathbf{x}$ is transmitted over the channel.

The  channel takes the codeword $\mathbf{x}$ as input and outputs a noisy version $\mathbf{y}=[y_1,\ldots,y_N]$, where the $y_i$ take real values. As mentioned before, having an information-theoretically defined channel model is not necessary, but if there is such a model, it is typically defined as a vector channel with transition probability density function (pdf) $W_N(\mathbf{y}|\mathbf{x})$. A widely used channel among researchers for code design is additive white Gaussian noise (AWGN) channel for which the output $y_i=x_i+w_i$ where $w_i$ is Gaussian random variable with zero mean and variance $\sigma^2$. For AWGN channel $W_N(\mathbf{y}|\mathbf{x})=\prod_{i=1}^N W(y_i|x_i)$, where $W(y|x)=\frac{1}{\sigma \sqrt{2\pi}} \mathrm{exp}^{ -\frac{{\left( y-x \right) }^2}{2\sigma^2}}$. The decoder network receives the channel output vector $\mathbf{y}$ and applies a decoding function $g_\phi (.) $  to give the decoded message word $\widehat{\mathbf{u}}=[\hat{u}_1,\ldots, \hat{u}_K]=g_\phi \left( \mathbf{y} \right)$ where the $\phi$ denotes the weights of the decoder neural network. The encoder and decoder networks together form an AE. The goal is to minimize the BLER or BER for different levels of impairment, e.g. SNR defined as $10\mathrm{log}_{10} 1/\sigma^2$ for the AWGN channel.

\section{ListAE}

\begin{figure}[!tb]
\begin{center}
\includegraphics[scale=.43,viewport=35 80 800 230,clip]{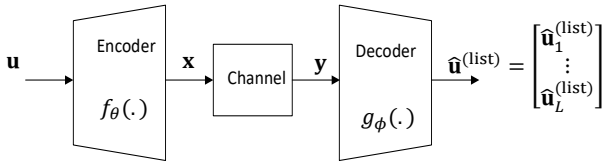}
\end{center}
\vspace*{-3mm}
\caption{A general listAE framework.}
\label{generalAE_list}
\end{figure}

Although designing new AE architectures can be a direction to improve the error correction performance, we choose to tackle the problem with a different approach. We posit that it may be too difficult for the decoder network to reliably decide which message word has been transmitted by only one guess. Therefore, we propose a framework which allows the decoder network to output a list of $L$ candidates. Figure \ref{generalAE_list} shows a general listAE with a list size $L$. A conventional AE is a special case of listAE with a list size of $L=1$. The concept of list decoding is well studied in the classical coding theory and, to a great deal, we have borrowed from that field. For example, successive cancellation list (SCL) of polar codes and its different variants have been well studied theoretically and also implemented for practical wireless communication systems \cite{sclTV}-\cite{CASCLKaiNiu}.


Since, in the testing phase, the decoder must output a single candidate $\hat{\mathbf{u}}$, there must be a selection process where a single candidate is chosen from the list. A GA decoder outputs $\widehat{\mathbf{u}}=\mathbf{u}$ if $\mathbf{u}$ is equal to one of the rows of $\widehat{\mathbf{u}}^{\left( \mathrm{list} \right)}$, otherwise it outputs a randomly chosen row of $\widehat{\mathbf{u}}^{\left( \mathrm{list} \right)}$. In other words,

\begin{align}
\widehat{\mathbf{u}} = \begin{cases}
\mathbf{u} &\text{if $\widehat{\mathbf{u}}_j^{ \left( \mathrm{list } \right)}=\mathbf{u}$ for any $j \in \{1,\ldots,L\} $}\\
\widehat{\mathbf{u}}_r^{ \left( \mathrm{list } \right)} &\text{otherwise}
\end{cases}
\label{GA_output_eq}
\end{align}
where $r$ is a random number chosen uniformly from $1$ to $L$. During the training phase, the value of each element of vectors in the output list $\widehat{\mathbf{u}}^{\left( \mathrm{list} \right)}$ is made to take a real number between zero and one, for example by passing through a Sigmoid activation. In the testing phase the outputs are rounded to the nearest integer to give binary values. It is also possible to select a single candidate by replacing the genie with CRC as it will be demonstrated later.

\subsection{Loss Function}

For a conventional AE, a number of loss functions, such as mean square error (MSE) and binary cross entropy (BCE), are more suitable for BER optimization. Although BER optimization indirectly optimizes the BLER, finding BLER-specific loss functions with rather efficient training complexity remains an open problem. 

With GA decoding, the performance metric to optimize is BLER which is calculated between $\mathbf{u}$  and $\widehat{\mathbf{u}}$ given by (\ref{GA_output_eq}). The challenge for defining a loss function which is tailored to the GA decoding of the listAE lies in how to mathematically model the genie operation. One may think of the genie operation as a processing block which takes the list of candidates as well as the transmitted message word and outputs a single candidate depending on the presence of the message word in the list. The condition for checking this presence involves rounding the candidate message words in the list to take binary values and then comparing them to the transmitted word. This operation a) introduces zero derivative in the back propagation, and b) additionally complicates it due to the comparisons. To tackle this problem, we propose a modified loss function that some how reflects how ``close" the output list is to the message word without involving the precise genie operation. The loss function should take small values when the message word is close to any candidate in the list and is defined as follows.



\begin{equation}
\mathrm{loss} \left( \widehat{\mathbf{u}}^{\left( \mathrm{list} \right)}, \mathbf{u} \right)= \mathrm{min}_{l \in \{1,\ldots,L \}} \rho \left( 
\widehat{\mathbf{u}}_l^{ \left( \mathrm{list } \right)}, \mathbf{u} \right)
\label{loss1}
\end{equation}

where $\rho$ is the average BCE loss function which takes two vectors $\widehat{\mathbf{x}}$ and $\mathbf{x}$ of length $K$.
\begin{eqnarray}
\rho \left(\hat{\mathbf{x}}, \mathbf{x}\right) \hspace{-8pt} &=& \frac{1}{K} \sum_{k=1}^K  \mathrm{bce} \left(\hat{x}_k, x_k \right) \nonumber \\ &=& -\frac{1}{K} \sum_{k=1}^K x_k \mathrm{log} \hat{x}_k +(1-x_k)\mathrm{log}(1-\hat{x}_k) 
\label{bce}
\end{eqnarray}

It is noteworthy that the function $\mathrm{min}(a,b)$  is in general non-differentiable as the derivative does not exist at points where $a=b$. Similarly, the derivative of the loss function does not exist at points where an equality holds between the input arguments. Such points happen with zero probability, so they do not cause any issue to the backpropagation of the gradients during the training, as we will see later.      

 With CA decoding and a $Z$ bit CRC generated by a polynomial $g(x)=g_0+g_1x+\ldots +g_Z x^Z$ a word of $K-Z$ bits is generated and is passed to the CRC calculator to generate $Z$ CRC bits. The CRC bits are appended to the end of the message word to give the length-$K$ vector $\mathbf{u}$ as the encoder input. At the decoder side, each candidate in the list is checked for passing CRC equations. Among the candidates which pass the CRC, one is randomly chosen as the final output of the decoder. 
 
To train listAE under CA decoding, we treat the CRC bits as information bits. In other words, the correlation between the bits of $\mathbf{u}$ is not taken into account to minimize the loss function. The reason is similar to those which leaded to employing the proposed loss function and avoiding the precise genie operation. Similarly, checking CRC involves binary Galois field operations which complicates the loss function and training. Therefore, we use the loss function given in (\ref{loss1}) for training both GA and CA decoding.

It is also noteworthy to mention that adding CRC to the message bits reduces the effective code rate by a factor of $Z/K$. To avoid rate reduction, one possible approach is to follow the SCL decoding of polar codes and assign a scalar metric $\rho_l, l=1,\ldots,L$, to each candidate in the list. Unfortunately we have not observed promising training results with this approach for listAE. However we think that this method is worth more investigation. 
%

\begin{figure*}[!t]
\begin{center}
\includegraphics[scale=.77,viewport=-10 108 635 307,clip]{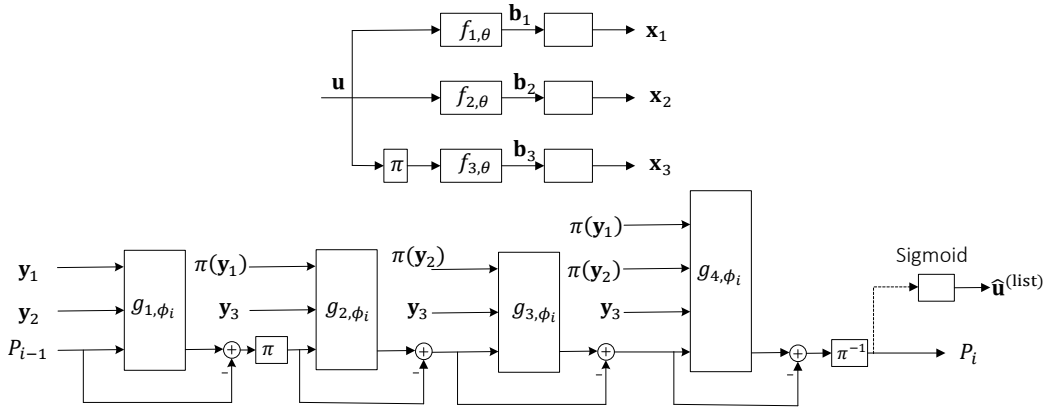}
\end{center}
\vspace*{-3mm}
\caption{Rate-$1/3$ IR-AE encoder and decoder at iteration $i$; The path of Sigmoid block only exists for the last iteration}
\label{IR3}
\end{figure*}

\subsection{An architecture for listAE: IR-AE}

In this section we present the IR-AE architecture. The encoder of IR-AE is essentially the same as Turbo-AE architecture and the decoder, too, relies on similar information exchange between the decoding blocks \cite{AE7turboae}. Roughly speaking a rate $1/n$ IR-AE uses $n$ encoding blocks which are applied to interleaved length-$K$ message words and give the length-$N$, $N=nK$, codeword $\mathbf{x}=\left[\mathbf{x}_1, \ldots, \mathbf{x}_n \right]$ after proper power normalization. Like Turbo-AE, IR-AE decoder consists of $I$ iterations. At iteration $i$, a series of decoding blocks which are serially concatenated take a certain subset of $\{ \mathbf{y}_1, \ldots, \mathbf{y}_n \}$ with applicable interleavers and a list matrix as input and output an updated list matrix. The same architecture is replicated in every iteration, but with independent learnable weights. If a decoding block takes a subset of $\{ \mathbf{y}_1, \ldots, \mathbf{y}_n \}$ consisting of $k$ vectors, i.e. $\{ \mathbf{y}_{i_1}, \ldots, \mathbf{y}_{i_k} \}$, we say that the decoding block is a rate-$1/k$ decoding block as one may associate it with an effective encoder which outputs the corresponding $k$ vectors $\{ \mathbf{x}_{i_1}, \ldots, \mathbf{x}_{i_k}\}$. An AE described as above is said to be an IR-AE if in an iteration each decoding block has a rate which is smaller or equal to the rate of the previous block. The heuristic motivation behind IR-AE architecture is to allow more powerful codes with smaller rates to attempt decoding the message word based on an improved list matrix given by previous weaker codes with higher rates. In this paper we mainly train and evaluate the performance of a rate-$1/3$ IR-AE. The detailed encoder and decoder architecture and training methodology is as follows.

\subsubsection{Rate-1/3 IR-AE} Fig. \ref{IR3} shows a rate-$1/3$ IR-AE. The encoder is identical to the rate-$1/3$ Turbo-AE encoder \cite{AE7turboae}. The output of the encoder is the length-$N=3K$ codeword with normalized power $\mathbf{x}=\left[\mathbf{x}_1, \ldots, \mathbf{x}_3 \right]$. The interleaver $\pi$ takes the length-$K$ word and outputs an interleaved word. The decoder consists of $I$ serially concatenated iterations. As can be seen, each iteration consists of four decoding blocks with rates $\left(1/2,1/2,1/2,1/3 \right)$, taking the inputs $\{ \mathbf{y}_1,\mathbf{y}_2\}$, $\{ \mathbf{y}_1,\mathbf{y}_3\}$, $\{ \mathbf{y}_2,\mathbf{y}_3\}$ and $\{ \mathbf{y}_1,\mathbf{y}_2\, \mathbf{y}_3\}$, respectively with applicable interleavers. An iteration $i$ also takes a list matrix $P_{i-1}$ of size $K \times L$ as an input and outputs a list matrix  $P_{i}$  for iteration  $i+1$. When taking a $K \times L$ matrix as input, the interleaver $\pi$ or deinterleaver $\pi^{-1}$ is applied on each column and generate a matrix of the same size. The list matrix $P_{I}$, output by iteration $I$, is additionally passed through a Sigmoid function and gives the output list of message word candidates. Our observation shows that the intermediate list matrices can be thought as the log-likelihood ratios (LLRs) of the message bits; as we move from the output of the first iteration to the next iterations, the BER resulting from the list matrices decreases. The interleavers are employed to mimic their role on enhancing the distance properties of the code by introducing long term memory \cite{AE7turboae}. At the decoder side, deinterleavers are applied similar to Turbo-AE and turbo codes. 
 

\subsubsection{Power normalization} \label{PN}
The output $\mathbf{b}=\left[\mathbf{b}_1, \ldots, \mathbf{b}_3 \right]$ is given to a power normalization block giving the codeword $\mathbf{x}=h\left(\mathbf{b} \right)$ to meet the power constraint requirements. Normalization can be done on code symbols, codewords or a batch of codewords \cite{AE7turboae}. In this paper we use the batch-wise normalization. With a batch size of $B$, and codewords of length $N$, there are $B.N$ code symbols. Each codeword is normalized as $\mathbf{x}=\frac{\mathbf{b}-\mu}{\gamma}$ where $\mu$ is the mean of the $B.N$ code symbols and $\gamma$ is the standard deviation of the $B.N$ code symbols. This normalization method places the set of $B.N$ code symbols in the batch on an $B.N$-dimensional sphere with radius $\sqrt{B.N}$. In the testing phase, $\mu$ and $\gamma$ can be pre-computed from a large batch and be directly used on a single message word. 
  

\subsubsection{Training methodology and hyper parameters}

\begin{table*}[t]
	\caption{Hyperparameters of the List IR-AE with $K=100$ and $N=300$}
	\centering	
	\begin{tabular}{ |p{4.5cm}|p{12cm}|  }
		\hline
					
		Encoder block $f_{k,\theta}(.)$ for $k =1,2,3$ & 5 layers 1D-CNN plus one linear layer, kernel size 5, CNN output channels 100\\
		\hline		
		First three decoding block $g_{k,\phi_{i}}$ for $k=1,2,3$ & 5 layers 1D-CNN plus one $(100,L)$ linear layer, the input and output channels of the CNN layers are $(L+2,100)$ and $(100,100)$ for the first and next layers respectively, kernel size 5\\
		\hline		
		Fourth decoding block $g_{4,\phi_{i}}$ & 5 layers 1D-CNN plus one $(100,L)$ linear layer, the input and output channels of the CNN layers are $(L+3,100)$ and $(100,100)$ for the first and next layers respectively, kernel size 5\\
		\hline
		
		 (learning rate, batch size $B$)  & $(.0001, 500)$ gradually changing to $(.000001,10000 )$\\
	 	\hline
		 Encoder and decoder training SNR & $1 \mathrm{dB}$ for encoder and $\left[-1.5,2 \right] \mathrm{dB}$ for decoder\\
	
	 	\hline
 Activation function & Elu for CNN layers and Linear for the linear layers\\
		 \hline		 
		 Optimizer, $(T_{enc}, T_{dec})$, Iterations  &  Adam, $(100,500)$, 6\\
		\hline
	\end{tabular}
	\label{hyper_param_table}
\end{table*}

Fully connected neural network (FCNN), CNN and RNN are natural choices to employ in Fig. \ref{IR3}. Our experiments show that FCNN is more difficult to train and provides inferior performance to CNN. RNN models such as Long-Short Term Memory (LSTM) and Gate Recurrent Unit (GRU) can bring global dependency, which in turn may improve the Euclidean distance properties of the code. However, according to our experiments, CNN models were easier to train and had superior performance to FCNN and RNN. Therefore, in this paper we mainly present the result for CNN-based IR-AE.


Table \ref{hyper_param_table} shows the details of the training and hyper parameters of our best IR-AE model. The model is trained for a maximum number of $500$ epochs. At each epoch, we train the encoder $T_{enc}$ times while freezing the weights of decoder, and then train the decoder $T_{dec}$ times while freezing the weights of encoder. This specific scheduled training was proposed in \cite{AE7turboae} to avoid getting stuck in local minima. With a batch size of $B$, for each training a set of $B$ randomly generated message words of length $K=100$ are generated and encoded by the encoder network. A set of $B$ noise vectors of length-$N$ are generated and added to the codewords corresponding to the message words. Following the methodology in \cite{AE7turboae}, a fixed SNR is used for training encoder while a range of SNR is used to train the decoder. For the latter, for each noise vector a SNR value is randomly picked from the range and is used to generate the noise vector.  Sufficiently large batch sizes with small learning rates are needed for fine tuning the model and are implemented according to \cite{AE6}. To decouple our investigation from interleaver design problem and for the sake of simplicity, in this paper a random interleaver is generated and fixed during the training and testing phase. There are existing works in the literature on interleaver desig of Turbo-AE which can be applicable to IR-AE. For instance, it is possible to design interleavers with uniform positional BER for Turbo-AE, which in turn may improve the overall performance. Interested readers are referred to \cite{ourglobecom} for more details.

\section{Experiment Results }
\begin{figure}[!tb]
\begin{center}
\includegraphics[scale=.52,viewport=20 12 765 440,clip]{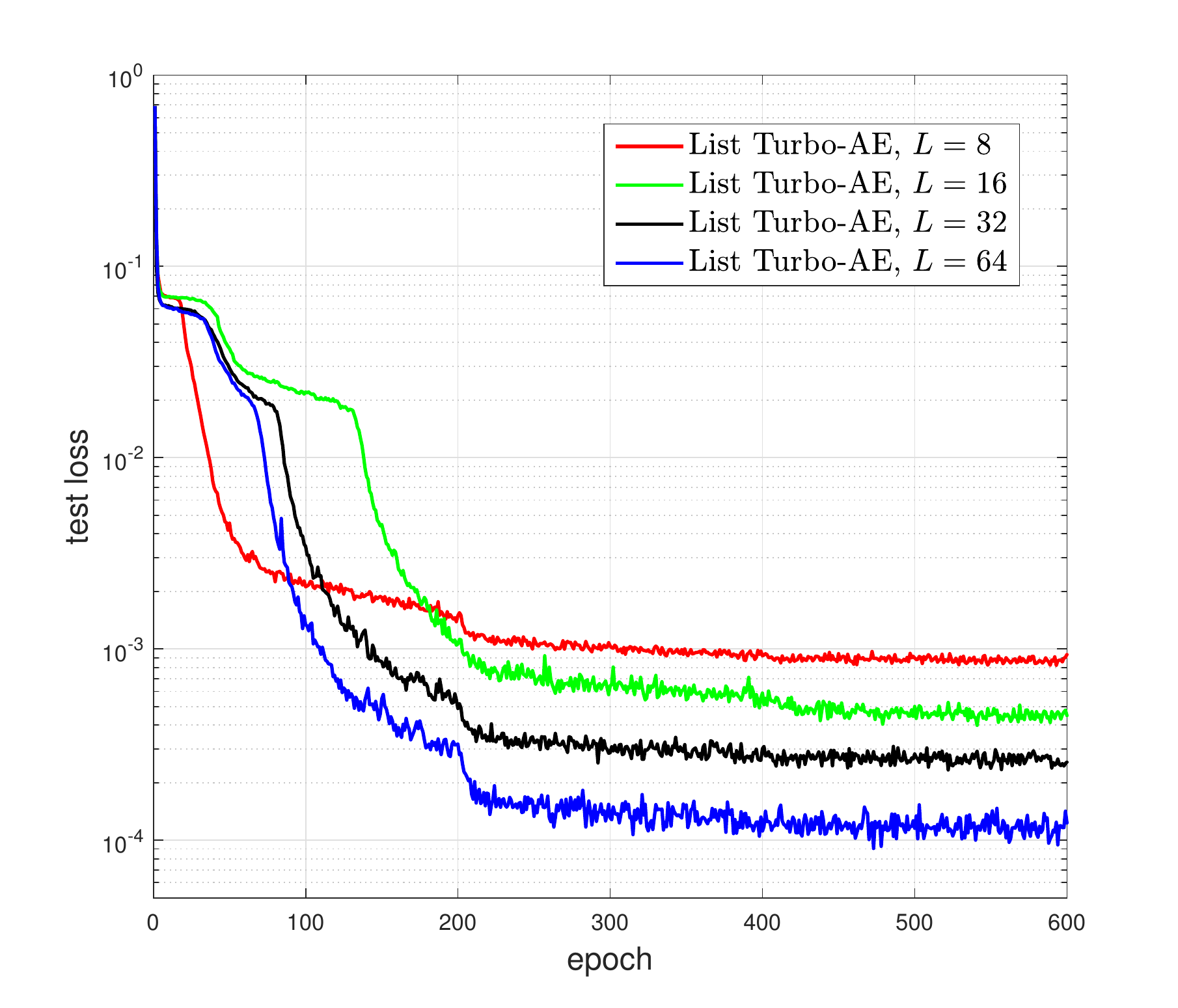}
\end{center}
\vspace*{-5mm}
\caption{Test loss trajectory for List Turbo-AE with different list sizes.}
\label{train_loss100_turboAE}
\end{figure}

\begin{figure}[!tb]
\begin{center}
\includegraphics[scale=.52,viewport=20 12 765 440,clip]{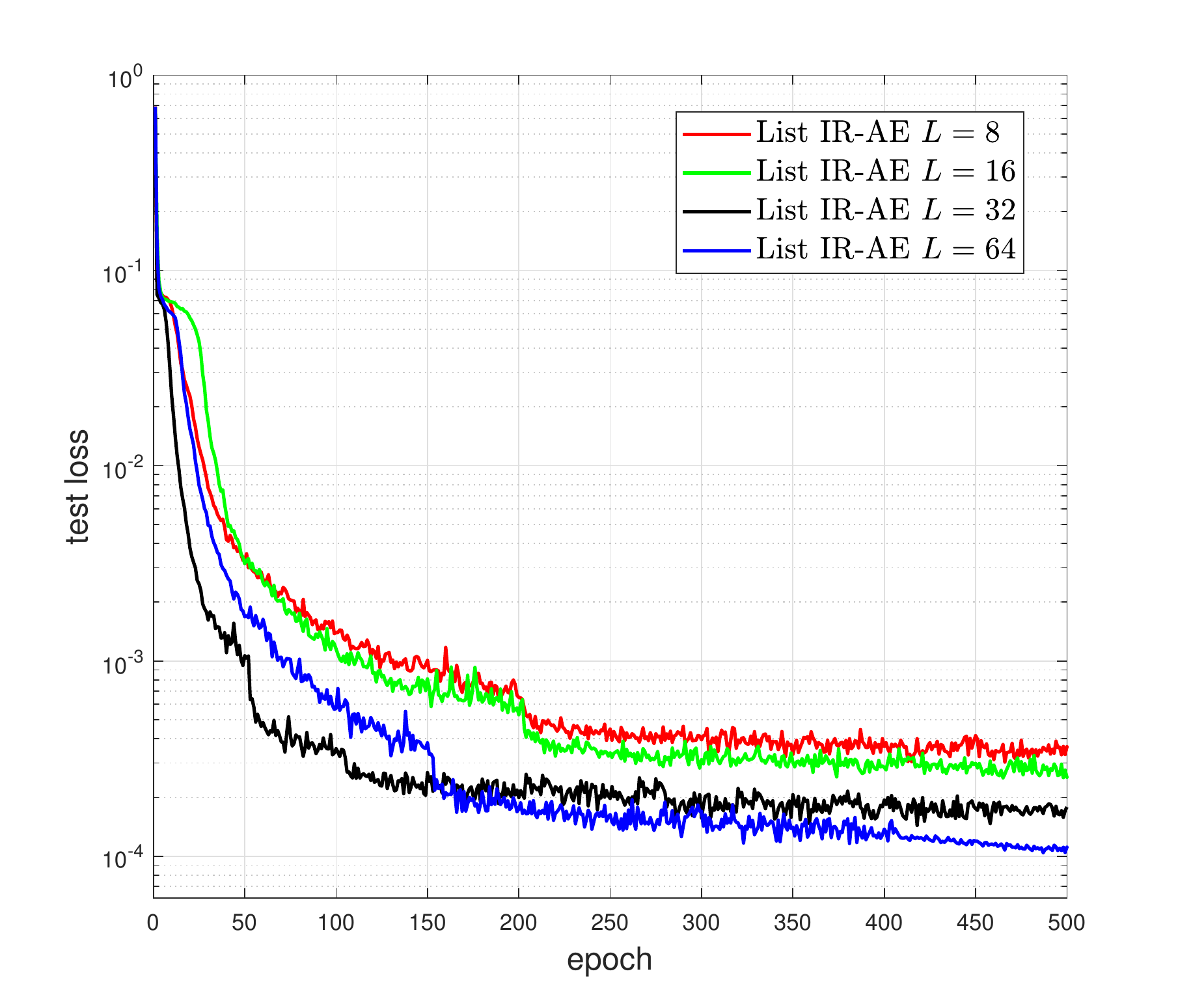}
\end{center}
\vspace*{-5mm}
\caption{Test loss trajectory for List IR-AE with different list sizes.}
\label{train_loss100_IR_AE}
\end{figure}

In this section, we present the performance results for the listAE with both Turbo-AE and IR-AE architectures and compare the results with the classical codes. For IR-AE, the hyper parameters are given in Table \ref{hyper_param_table}. The parameters for Turbo-AE are the same as the relevant blocks of the IR-AE, i.e. the first two decoding blocks. Figures \ref{train_loss100_turboAE} and \ref{train_loss100_IR_AE} show the resulting test loss for List Turbo-AE and List IR-AE respectively. For each epoch, after the model is trained, the test loss is calculated for a new set of training examples generated for the training SNR of $1$ $\mathrm{dB}$. An interesting observation is that Turbo-AE appears to be more sensitive to the list size. When we change the list size from 8 to 64, the converged value of loss drops almost one order of magnitude for Turbo-AE while the change is quite smaller for IR-AE. This would suggest an advantage of the latter over the former at smaller list sizes. Also, as expected, the test loss generally decreases from a smaller list size to a larger one. However, this trend does not hold at every epoch, which is probably because the optimizer needs to observe more data/epochs to train for a larger list size due to the increased model size. 


%

\begin{figure}[!tb]
\begin{center}
\includegraphics[scale=.52,viewport=20 12 765 440,clip]{ 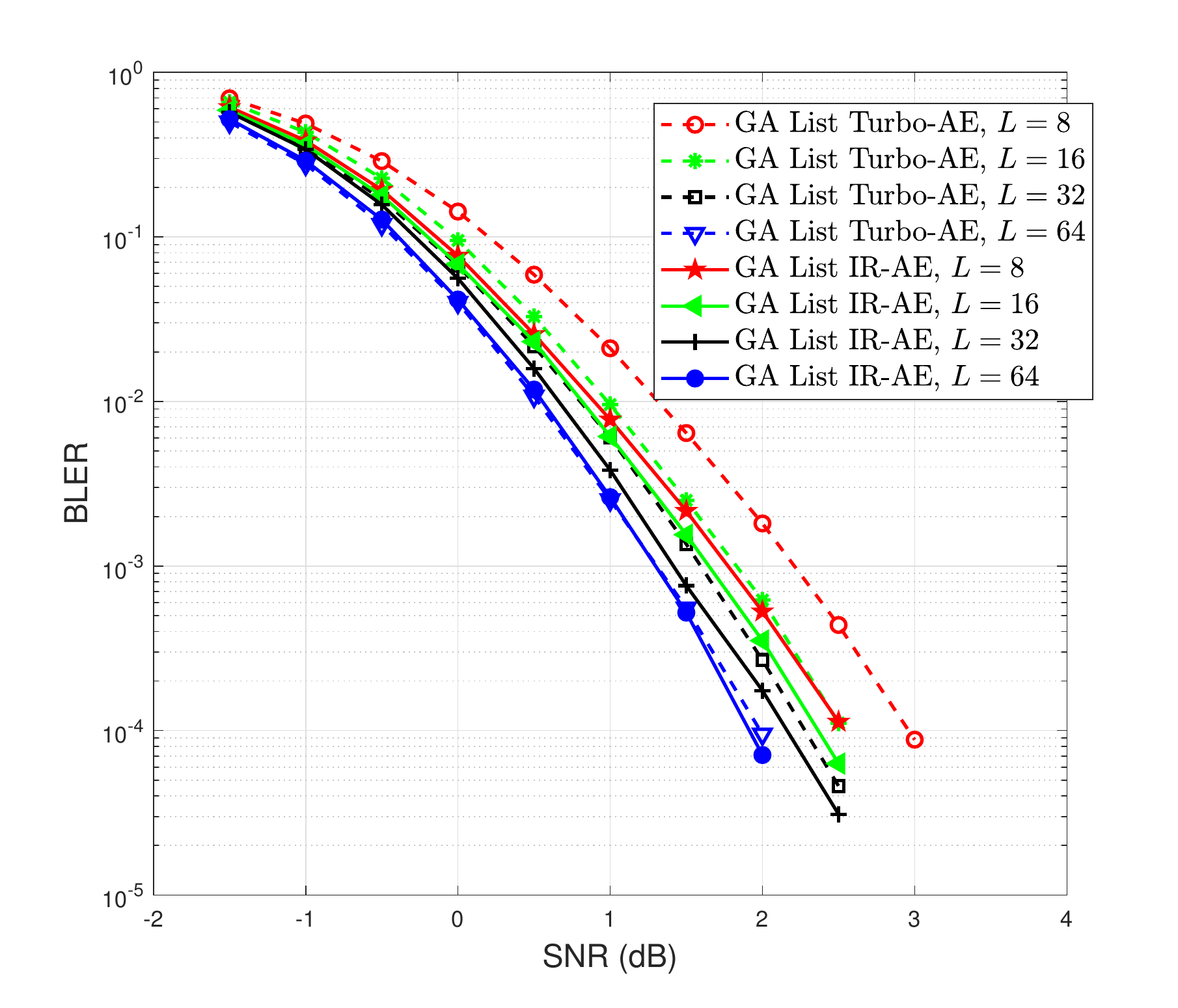}
\end{center}
\vspace*{-5mm}
\caption{Performance comparison of List IR-AE and List Turbo-AE for $K=100$ and $N=300$}
\label{ER100_onlyGA}
\end{figure}

Fig. \ref{ER100_onlyGA} demonstrates the BLER of the List Turbo-AE and List IR-AE under GA decoding for different list sizes. As shown, the performance for each architecture follows the trend given by the test loss trajectory. The trajectory also implies that for larger list sizes the IR-AE and Turbo-AE have similar performances, whereas for smaller list sizes IR-AE outperforms Turbo-AE. The improvement comes at the price of more than twice decoding network size as the Turbo-AE. For the remainder of the paper we focus on List IR-AE.

\begin{figure}[!tb]
\begin{center}
\includegraphics[scale=.52,viewport=20 12 765 440,clip]{ 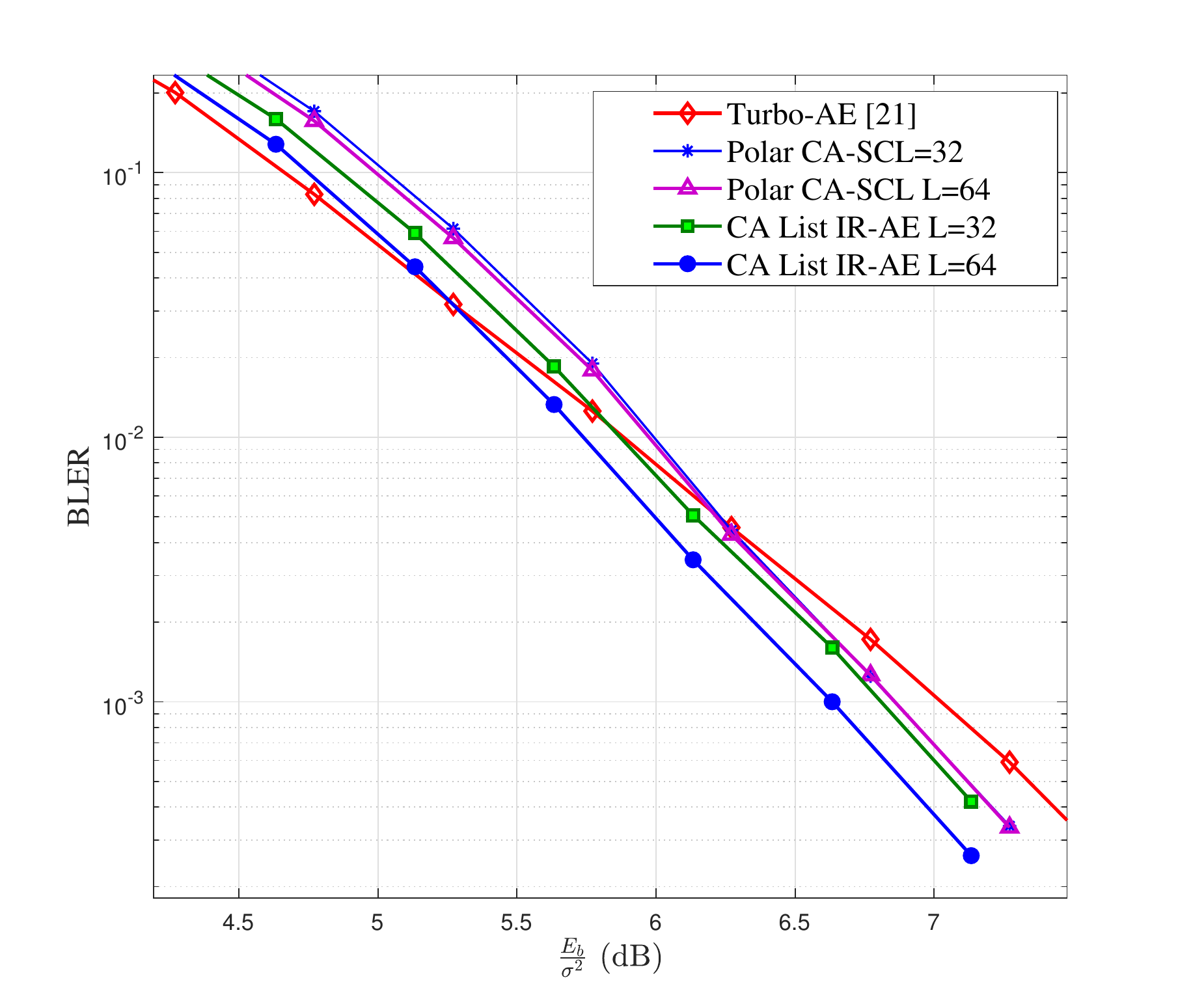}
\end{center}
\vspace*{-5mm}
\caption{Performance comparison between List IR-AE, Turbo-AE and polar code}
\label{ER100_final}
\end{figure}

Next, we evaluate the performance of the List IR-AE under CA decoding and compare it to classical codes and Turbo-AE \cite{AE7turboae}. The code dimensions for the Turbo-AE and polar codes are $(N=300,K=100)$. The polar codes are designed according to 3GPP NR reliability sequence and rate matching. For IR-AE a length-8 CRC generated by polynomial $g(x)=1+x^2+x^4+x^6 + x^7 +x^{8}$ is appended to the $K=92$ message bits before encoding. To have a fair comparison due to slight rate reduction by CRC, we look at $E_b/\sigma^2$ instead of SNR for a code of rate $R$ where $E_b/\sigma^2 = \mathrm{SNR}-10\mathrm{log}_{10} (R)$. The result of the comparison is shown in Fig. \ref{ER100_final} only for large list sizes for the best performance. As can be seen, List IR-AE with a list size of 64, outperforms Turbo-AE and the polar code at BLERs smaller than $0.03$. At high SNRs, the coding gain can be as large as $0.5$ $\mathrm{dB}$ and $0.3$ $\mathrm{dB}$ over Turbo-AE and polar code, respectively.

\section{Conclusion}

In this paper, we presented listAE motivated by the goal of bringing the list decoding in classical coding theory into the area of deep learning based channel encoder and decoder design. To the best of our knowledge, this is the first work that introduces list decoding in the field of deep learning based channel code design. Our simulation results show performance gain over Turbo-AE and polar code which is promising and calls for future research. This work sets the stage to design AEs which can compete or outperform classical codes. For future directions, we note that the current choice of the loss function was somewhat heuristic and intended to optimize the performance under GA list decoding. Finding better loss functions to reflect the performance under CA list decoding is an interesting future direction.


\begin{thebibliography}{9}
\bibitem{Shannon}
C. E. Shannon, ``A mathematical theory of communication," \emph{ The Bell System Technical Journal}, vol. 27, no. 3, pp. 379-423, 1948.

\bibitem{TurboCode}
C. Berrou, A. Glavieux and P. Thitimajshima, ``Near Shannon limit error-correcting coding and decoding: Turbo-codes. 1," \emph{ in Proc. IEEE International Conference on Communications (ICC)}, pp. 1064-1070, vol.2, 1993.

\bibitem{LDPCGallager}
R. G. Gallager, \emph{Low-Density Parity-Check Codes.} Cambridge, MA: MIT Press, 1963


\bibitem{PolarArikan}
E. Arikan, ``Channel polarization: A method for constructing capacity-achieving codes for symmetric binary-input memoryless channels," \emph{IEEE Trans. Inf. Theory}, vol. 55, no. 7, pp. 3051–3073, Jul. 2009

\bibitem{decoder1}
A. Hamalainen and J. Henriksson, ``A recurrent neural decoder for convolutional codes,” in \emph{Proc. IEEE International Conference on Communications (ICC)}, vol. 2, pp. 1305–1309, 1999.

\bibitem{decoder2}
T. Gruber, S. Cammerer, J. Hoydis, and S. t. Brink, ``On deep learning based channel decoding,” in \emph{ Proc. 51st Annual Conference on Information Sciences and Systems (CISS)}, 2017, pp. 1–6.

\bibitem{decoder3}
H. Kim, Y. Jiang, R. Rana, S. Kannan, S. Oh, and P. Viswanath,``Communication algorithms via deep learning,” in \emph{ Proc. international conference on representation learning (ICLR)}, 2018.

\bibitem{decoder4}
E. Nachmani, Y. Be’ery, and D. Burshtein, ``Learning to decode linear codes using deep learning,” in \emph{Proc. 54th Annual Allerton Conference on Communication, Control, and Computing (Allerton)}, pp. 341–346, 2016

\bibitem{decoder5}
S. Cammerer, T. Gruber, J. Hoydis, and S. ten Brink, ``Scaling deep learning-based decoding of polar codes via partitioning,” in \emph{ Proc. IEEE Global Communications Conference (GLOBECOM)}, pp. 1–6, 2017

\bibitem{Denoising}
A. Makhzani and B. Frey, ``K-sparse autoencoders," \emph{arXiv preprint arXiv:1312.5663}, 2013.


\bibitem{generativemodel1}
D. P. Kingma and M. Welling, “Auto-encoding variational bayes," \emph{arXiv preprint arXiv:1312.6114}, 2013.


\bibitem{generativemodel2}
X. Chen, Y. Duan, R. Houthooft, J. Schulman, I. Sutskever, and P. Abbeel, ``Infogan: Interpretable representation learning by information maximizing generative adversarial nets," in \emph{Advances in neural information processing systems}, 2016, pp. 2172–2180.


\bibitem{representationlearning1}
P. Vincent, H. Larochelle, Y. Bengio, and P.-A. Manzagol, ``Extracting and composing robust features with denoising autoencoders,” in \emph{Proc. of the 25th international conference on Machine learning}, ACM, 2008, pp. 1096–1103.


\bibitem{representationlearning2}
A. Krizhevsky and G. E. Hinton, ``Using very deep autoencoders for content-based image retrieval", in \emph{ESANN}, 2011.

\bibitem{AE1}
T. O'Shea and J. Hoydis, ``An introduction to deep learning for the physical layer,” \emph{ IEEE Trans. Cognitive Communications and Networking}, vol. 3, no. 4, pp. 563–575, 2017.

\bibitem{AE2}
Z. Qin, H. Ye, G. Y. Li, and B.-H. F. Juang, ``Deep learning in physical layer communications,” \emph{IEEE Wireless Communications}, vol. 26, no. 2, pp. 93–99, 2019.

\bibitem{AE3}
H. Ye, L. Liang, and G. Y. Li, ``Circular convolutional auto-encoder for channel coding,” in \emph{Proc.  20th IEEE International Workshop on Signal
Processing Advances in Wireless Communications (SPAWC)}, 2019, pp. 1–5.

\bibitem{AE4} 
Y. Jiang, H. Kim, H. Asnani, S. Kannan, S. Oh, and P. Viswanath, ``Learn codes: Inventing low-latency codes via recurrent neural networks,” in \emph{Proc. IEEE International Conference on Communications (ICC)}, 2019, pp. 1–7.


\bibitem{AE5} 
A. Felix, S. Cammerer, S. D¨orner, J. Hoydis, and S. Ten Brink, ``Ofdm-autoencoder for end-to-end learning of communications systems,” in \emph{Proc. 19th IEEE  International Workshop on Signal Processing Advances in Wireless Communications (SPAWC)}, 2018, pp. 1–5.

\bibitem{AE6}
M. V. Jamali, H. Saber, H. Hatami and J. H. Bae, ``ProductAE: Towards training larger channel codes based on neural product codes," \emph{arXiv preprint arXiv:2110.04466}, 2021.

\bibitem{AE7turboae}
Y. Jiang, H. Kim, H. Asnani, S. Kannan, S. Oh, P. Viswanath, ``Turbo Autoencoder: Deep learning based channel codes for point-to-point communication channels", \emph{in Proc. 33rd Conf on Neural Information Processing Systems (NeurIPS 2019)}. 

\bibitem{sclTV}
I. Tal and A. Vardy, ``List Decoding of Polar Codes," in \emph{IEEE  Trans. Inf. Theory}, vol. 61, no. 5, pp. 2213-2226, May 2015.

\bibitem{CASCLKaiNiu}
K. Niu and K. Chen, ``CRC-Aided Decoding of Polar Codes," in \emph{IEEE Communications Letters}, vol. 16, no. 10, pp. 1668-1671, October 2012

\bibitem{ourglobecom}
H. Yildiz, H. Hatami, H. Saber, Y. Cheng and J. Bae, ``Interleaver Design and Pairwise Codeword Distance Distribution Enhancement for Turbo Autoencoder," in \emph{ Proc. IEEE Global Communications Conference (GLOBECOM) 2021}


\end{thebibliography}
\end{document}